\documentclass[12pt,preprint]{aastex}

\newcommand{\be}{\begin{equation}}
\newcommand{\ee}{\end{equation}}

\shorttitle{SPM4 Catalog}

\begin{document}

\title{The Southern Proper Motion Program IV. \\
    The SPM4 Catalog}


\author{Terrence M.~Girard, William F.~van Altena}
\affil{Astronomy Department, Yale University, P.O. Box 208101,
       New Haven, CT 06520-8101}
\email{terry.girard@yale.edu, william.vanaltena@yale.edu}

\author{Norbert Zacharias}
\affil{United States Naval Observatory, Washington, DC 20392, USA}
\email{nz@usno.navy.mil}

\author{Katherine Vieira\footnote{Current address: Centro de Investigaciones 
de Astronomia, Apartado Postal 264, M\'{e}rida 5101-A, Venezuela},
 Dana I.~Casetti-Dinescu}
\affil{Astronomy Department, Yale University, P.O. Box 208101,
       New Haven, CT 06520-8101}
\email{kvieira@cida.ve, dana.casetti@yale.edu}

\author{Danilo Castillo}
\affil{Department of Science Operations, San Pedro de Atacama, 
II Region-ALMA, Chile}
\email{daniloeros@hotmail.com}

\author{David Herrera}
\affil{National Optical Astronomy Observatory, 950 North Cherry Av.,
Tucson, AZ 85719}
\email{dherrera@noao.edu}

\author{Young Sun Lee, Timothy C.~Beers}
\affil{Department of Physics and Astronomy and the Joint Institute for Nuclear
Astrophysics, Michigan State University, East Lansing, MI 48824}
\email{lee@pa.msu.edu, beers@pa.msu.edu}

\author{David G.~Monet}
\affil{US Naval Observatory, Flagstaff Station, P.O. Box 1149, 
Flagstaff, AZ 86002}
\email{dgm@nofs.navy.mil}

\author{Carlos E.~L\'{o}pez}
\affil{Universidad Nacional de San Juan, Observatorio Astron\'{o}mico 
``F\'{e}lix Aguilar'' and Yale Southern Obs., 
Av. Benav\'{i}dez 8175 Oeste, Chimbas, 5413 San Juan, Argentina}
\email{cel\_2018@yahoo.com.ar}

\begin{abstract}

We present the fourth installment of the Yale/San Juan Southern Proper
Motion Catalog, SPM4.  
The SPM4 contains absolute proper motions, celestial coordinates,
and $B,V$ photometry for over 103 million stars and galaxies between the south
celestial pole and -20$^{\circ}$ declination.  
The catalog is roughly
complete to $V$=17.5 and is based on photographic and CCD observations
taken with the Yale Southern Observatory's double-astrograph at Cesco
Observatory in El Leoncito, Argentina.
The proper-motion precision, for well-measured stars, is estimated to
be 2 to 3 mas~yr$^{-1}$, depending on the type of second-epoch material.
At the bright end, proper motions are on the
International Celestial Reference System by way of Hipparcos Catalog stars,
while the faint end is anchored to the inertial system using external
galaxies.
Systematic uncertainties in the absolute proper motions are on the order
of 1 mas~yr$^{-1}$.

\end{abstract}


\section{Introduction}


The Yale/San Juan Southern Proper Motion program (SPM) is a decades-long
undertaking to determine absolute proper motions of southern-sky
stars for the purpose of better understanding our Galaxy's 
structure and kinematics.
The SPM was founded as the southern-sky counterpart to the
Lick Northern Proper Motion program (see Klemola et al.~1987 and Hanson
et al.~2004) and has benefited greatly from lessons learned in that venture.
The original impetus behind the SPM and its planning and organization are
described in detail by van Altena et al.~(1990, 1994).

Over the years, the SPM has grown to encompass a diverse set of projects ranging
from targeted, optimal-accuracy studies to more general, large-area
surveys.
Among these are 
the numerous determinations of absolute proper motions
of Milky Way globular clusters summarized by Casetti-Dinescu et al.~(2010),
with the SPM program being responsible for over half of all clusters,
in either hemisphere, with well-measured absolute proper motions;
a proper-motion study of the Magellanic Clouds (Vieira et al.~2010), in which
the large-format, overlapping SPM material allows for a unique
measure of the motion of the SMC relative to the LMC;
a determination of the absolute motion of the Sagittarius dwarf galaxy
(Dinescu et al.~2005a) and 
of the Canis Major dwarf galaxy (Dinescu et al.~2005b);
a determination of the velocity shear of the Milky Way thick disk
(Girard et al.~2006) using a sample of SPM stars at the South Galactic Pole;
a further investigation of thick-disk kinematics using a larger sample of SPM 
stars with radial velocity measures (Casetti-Dinescu et al.~2011);
and an exploration of possible open cluster remnants via candidate star 
proper motions (Carraro et al.~2005).
More utilitarian uses of SPM material have included
confirmation of the precursor to SN 1987a (Girard et al.~1988);
identifying the optical counterpart to non-optical discoveries, e.g.
GRO J1655-40 (Bailyn et al.~1995);
helping to establish the link between the Hipparcos system and that of
the ICRS (Platais et al.~1998b);
and providing positional input to other proper-motion projects such
as the UCAC3 (Zacharias et al.~2010).
The raw catalog products of the SPM are a series of astrometric and
photometric catalogs of successively expanded sky coverage and
completeness.
The fourth catalog in this series, the SPM4, is presented here.

In what follows, we provide a brief historical background on the
SPM in Sect.~2 and a description of the observations and measurements in
Sect.~3.
Details of the catalog construction, i.e., the
astrometric and photometric reductions, are given in Sect.~4 while
Sect.~5 takes a look at the properties of the resulting catalog.
Sect.~6 provides an overview of previous SPM catalogs, to help clarify the
relevant differences for the potential user.
Finally, an overall summary is given in Sect.~7 along with information
as to obtaining or accessing the catalog.

\section{Historical Background}

In 1960, the Yale-Columbia Southern Observatory, under the leadership
of Dirk Brouwer of Yale University and Jan Schilt of Columbia
University, was granted \$750,000 by the Ford Foundation with the goal of
building an observatory to determine the accurate
positions and apparent motions of the stars to study the structure of
our Milky Way Galaxy from the Southern Hemisphere. 
Following a survey
of potential sites in Australia, Chile and Argentina, an observatory
was built at El Leoncito, Argentina. 
The observatory was jointly
operated by the University of Cuyo's Observatorio
Astron\'{o}mico ``Felix Aguilar'' (OAFA) in
San Juan, under the leadership of Carlos U. Cesco, Jose Augusto
L\'{o}pez and the Yale-Columbia Southern Observatory, Inc., (YCSO). 

The first survey of the Southern sky, now known as the Yale/San Juan
Southern Proper Motion survey, or the SPM, was made between the years
1965 and 1979 under the direction of Adriaan J. Wesselink, Pierre
Demarque and William van Altena at Yale following the death of Dirk
Brouwer in 1966. 
Financial support for the survey was obtained through
a series of grants from the US National Science Foundation. 
Due to
changing research priorities at Columbia University, Columbia withdrew
from the Corporation in 1974 and the name was officially
changed to Yale Southern Observatory, Inc.~(YSO) on January 23, 1975. 
At that time William van Altena joined Yale and assumed the direction of
the YSO; he continues to lead the organization and direct the SPM.
About two years earlier, the University of Cuyo had split into several
regionally based units. 
The one based in San Juan became known as the
Universidad Nacional de San Juan (UNSJ) and it assumed the
administration of the OAFA and partnership with the YCSO and then the
YSO. 
After Carlos E. L\'{o}pez arrived at the OAFA in 1980 he assumed
responsibility for the YSO operations in Argentina and continues in
that role. Subsequent directors of the OAFA and Rectors of
the UNSJ made substantial contributions to the progress of the SPM.
In 1990 the El Leoncito Observatory was renamed the Dr.~Carlos 
U.~Cesco Observatory on the occasion of the 25th anniversary of the
beginning of observations in honor of Dr.~Cesco's many
contributions to the founding and operation of the Observatory.

\section{Observations and Plate Measurement}

SPM observations have been carried out using the 51-cm double astrograph of
the Cesco Observatory in El Leoncito, Argentina.
The double astrograph is essentially two telescopes sharing a common 
structural assembly and mount.
One of the two objective lenses is optimized for the visual passband, the
other for blue.
The photographic portion of the SPM survey was taken on quarter-inch thick
17x17 inch plates; 103a-G emulsion with OG515 filter for the yellow lens and
unfiltered 103a-O emulsion for the blue lens.
The first-epoch survey, taken from 1965 to 1979, is entirely photographic.
The second-epoch survey is approximately 1/3 photographic (taken from 1988
to 1998) and 2/3 CCD-based (taken from 2004 through 2008).  
Figure 1 shows the survey coverage differentiated by second-epoch
detector type.
The survey 
consists of fields at 5-degree centers in declination and varying offsets 
along right ascension, but always less than or equal to 5 degrees.  Since
each photographic plate covers a 6.3 x 6.3 degree area of sky, there is
significant overlap in the photographic portion of the survey.  Also, each
field has a pair of blue and yellow passband plates taken (typically)
simultaneously with the double-astrograph.  For a small fraction of the
fields, plates were repeated within the same ``epoch''.  Each photographic
observation consisted of two offset exposures, one 2 hours in duration, the
other 2 minutes.  
All photographic exposures were centered on the meridian.
An objective wire grating, with grating constant $\Delta$m=3.85, was used
to produce measurable diffracted images of the brighter stars.  In this
manner, the effective dynamic range of the plates was greatly increased,
allowing bright Hipparcos-magnitude stars to be linked to external galaxies
on the same plate.  
In addition to extending the dynamic range, the grating images also
provide a means for calibration and correction
of each plate's magnitude equation.
A more thorough description of the plate material and
the various image systems is given by Girard et al.~(1998).

The SPM plates were scanned with the Precision Measuring Machine (PMM) at the
US Naval Observatory's Flagstaff Station (NOFS).  
(See Monet et al.~2003 for a description of the PMM.)
The raw pixel data were
saved and later analyzed at the US Naval Observatory in Washington (USNO),
to obtain centers and photometric indices for all detectable images.

Beginning in 2000, CCD cameras were installed on the double astrograph in
order to complete the SPM second-epoch survey.  (Photographic plates with
the 103 emulsion were no longer being produced.)  Two cameras were installed,
a 4K x 4K PixelVision (PV) camera (15-micron pixels) in the focal plane of
the yellow lens, and an Apogee 1K x 1K (24-micron pixels) camera behind the
blue lens.  The blue camera was later replaced by an Apogee Alta 2K x 2K 
(12-micron pixels) CCD camera.  
Custom Scientific V and B band filters were incorporated into the
PV and Apogee systems, respectively.
This results in a passband response for each camera system that roughly
approximates its first-epoch photographic counterpart.
As with the plates, the objective grating
was used during CCD exposures.
During CCD observations, the grating is oriented such that the dispersion
direction is 45$^{\circ}$ from North and, hence, 45$^{\circ}$ with respect
to the columns and rows of the CCD.
This ensures, for bright stars, that the grating images will not be spoiled 
by column bleeding of the saturated central image.
Exposure times were 120-s, reaching the same
depth as the first-epoch plates.  
A two-fold spatial overlap of frames based on
the PV's 0.93 x 0.93 degree FOV was initiated for all SPM fields lacking
second-epoch plate material.  Eventually, when it was found that a single
CCD exposure was superior to the multiple first-epoch plate material in
terms of astrometric precision, the two-fold coverage was changed to single
coverage with the PV frames.  The yellow lens' PV data were used for both
astrometry and photometry.  The blue lens' Apogee data were used only in
the photometric reductions.

\section{SPM4 Construction}

\subsection{Input Catalog}

The astrometric reductions, of both the photographic and CCD data, made
use of an input ``master'' catalog that was necessary to properly identify
the various multiple images, 
i.e., diffraction grating orders and, in the case of
the plates, multiple exposures.  This master catalog was constructed by
combining the following external catalogs in the specified order:

\noindent
 1 = Hipparcos (Perryman et al.~1997)\\
 2 = Tycho2 (H{\o}g et al.~2000)\\
 3 = UCAC2 (Zacharias et al.~2004)\\
 4 = 2MASS psc (Skrutskie et al.~2006)\\
 5 = 2MASS xsc (Skrutskie et al.~2006, largely (but not 
entirely) galaxies) \\
 6 = LEDA      (confirmed galaxies, Paturel et al.~2005, A\&A 430, 751) \\
 7 = QSO       (Veron-Cetty \& Veron 2006, A\&A 455, 773) \\

Objects appearing in multiple catalogs were found by positional coincidence
and reconciled by adopting the position of the higher ranked source (Hipparcos
being considered best).  This master input catalog was then used to identify
all measurable images within the list of detections in the SPM plate and
CCD data.  Thus, an object that does not appear in any of these input
catalogs, cannot appear in the SPM4 catalog.  The completeness of the SPM4
is the product of the completeness of these input catalogs and the magnitude
limits and resolving limits (i.e., ability to center crowded/blended images)
of the SPM material.

There are 670 SPM field centers at declination -20 degrees and southward.
The SPM4 is comprised of 660 of these fields.  There are nine -20-deg fields
for which first-epoch plates were never taken and one -20-deg field for which
the first-epoch plates were mistakenly not measured.  Thus, the northern
boundary of the SPM4 sky coverage contains a small number of ``notches''
at which the northernmost stars are at approximately -21.9 degrees instead
of -20 degrees.

An input master catalog, as described above, was constructed from cutouts
of the external source catalogs for each of the 660 SPM fields included
in the SPM4.  Within a single field, each object was assigned a ``master''
ID number that was simply the running number corresponding to the order of
that object in the field's cumulative list.  Thus, Hipparcos stars would be
assigned the lowest numbers, increasing through Tycho2 stars, UCAC2 stars,
etc.  Combining the three digit field number with the seven digit master
catalog number within a field yields a star's overall SPM4 ID number.
Since many stars in the substantial overlap of neighboring fields would
possess multiple IDs, the ID from the lowest numbered field was adopted as
the unique SPM4 identifier.


\subsection{Astrometric Reduction}

All SPM plates were scanned with the PMM at NOFS.  The raw pixel data from
the scans were stored and sent to USNO for analysis.  The existing StarScan
pipeline (Zacharias et al.~2008) was heavily modified by USNO staff to
accommodate the SPM pixel data.  The overall process included a conversion of
the PMM transmission values into density values, smoothing the data for the
purposes of image detection and background fitting, and then fitting the
unsmoothed 2-D density profiles with an azimuthally symmetric exponential
function.  (Tests using an elliptical exponential function showed no
improvement over the azimuthally symmetric one, even for the slightly
elongated 1st and 2nd-order diffraction images.)  The derived image positions
on each of the 884 PMM CCD footprints required to cover an SPM plate were then
transformed into a single global coordinate system using information from the
overlap regions of adjacent footprints and the laser interferometric metrology
of the footprint centers. 
As the resulting astrometry from all first-epoch
SPM plates was included in the construction of the UCAC3 catalog, further
discussion of the PMM data analysis can be found in Zacharias et al.~(2010).
The USNO-derived centers and image parameters for all detections on the SPM
plates, both first and second-epoch plates, were then provided to the Yale SPM
team for subsequent reduction.

The SPM CCD frames are corrected for bias, dark (in the case of the Apogee
frames, dark current is negligible in the PV), and flatfielding.  SExtractor
(Bertin \& Arnouts 1996)
is used to identify detections, give aperture photometry, and provide
preliminary $(x,y)$ centers.  Final $(x,y)$ centers are derived by fitting
two-dimensional elliptical Gaussian functions to the image intensities.
See Casetti-Dinescu et al.~(2007) for further details of the astrometric
reduction procedures used with the PixelVision camera data.

In general, similar techniques to those used in constructing previous
versions of SPM catalogs were used to build the SPM4.  (See Girard et 
al.~1998, Platais et al.~1998a, Girard et al.~2004.)  Stars for which both
the central-order exposure and first-order grating-image pair are
measurable were used to derive and correct each plate's magnitude equation
individually.  Following the procedures developed for the SPM1 and SPM2
catalogs, all extended sources were given magnitude corrections corresponding
to their measured magnitude shifted by -0.7, an empirically determined
correction that compensates for the difference between a plate's star-image and 
galaxy-image magnitude equations.

In the case of the CCD image centers, there were systematic offsets
detected between the position of the central image and the mean of the
positions of the grating-order pairs.  However, these did not follow the
behavior expected for magnitude equation or charge transfer efficiency
effects.  Therefore, this offset was corrected as such, a simple offset
between the image order systems, instead of as a magnitude equation.
(See Casetti-Dinescu et al.~2007.)

Measures from all exposures and all grating-image systems were transformed
to a single system for each plate and for each CCD frame.  The CCD $(x,y)$
positions from the PV frames were also corrected for a fixed-pattern 
geometric distortion believed to be associated with the filter.  
Figure 2 shows the distortion pattern.
The correction ``mask'' was built up
from residuals of hundreds of frames, at different pointings, reduced into
UCAC2 coordinates.  The corrected CCD $(x,y)$ positions were then transformed
onto the system of the UCAC2 to facilitate pasting together the roughly 50
to 100 frames (depending on whether it had two-fold or single coverage) that
comprise a 6 deg x 6 deg SPM field.  An overlap method is employed to
perform this task, using Tycho2 stars as an external reference system to
ensure that systematics from the individual overlap solutions do not
accumulate.  In this manner, an artificial ``pseudo-plate'' is built up from
CCD frames.  This pseudo-plate can then be treated the same as a real
second-epoch plate.

In previous versions of SPM catalogs, first- and second-epoch plate pairs
were combined to yield relative proper motions per plate pair.  These
were then corrected to absolute proper motions using external galaxies in
the case of SPM1 and SPM2, or Hipparcos star proper motions in the case of
SPM3.  For the SPM4, instead of combining plate pairs, we have decided to
construct the best possible position catalogs at first and at second epoch,
over the entire coverage area.  This is accomplished by dividing the
plates into three groups -- first-epoch plates, second-epoch plates, and
second-epoch pseudo-plates -- and combining positional data within each 
group using a plate-overlap strategy as follows.

Within each of the plate groups, all plates are pushed through the software
pipeline that performs the preliminary reductions described above.  This
pipeline combines multiple images of the same star (short/long exposure and
diffraction orders), corrects the positions for magnitude equation, and then
models these $(x,y)$ positions into $(\alpha,\delta)$ 
from a subset of the Tycho2 catalog
adjusted to the epoch of the plate.  (The subset is roughly half the Tycho2
stars, those of better quality.)  The plate model consists of classical
5th-order distortion terms plus a general 3rd-order polynomial.  Uncertainties
as a function of magnitude are derived from the scatter in position
of stars with multiple images.

With each plate having been reduced into $(\alpha,\delta)$ 
on the system of Tycho2,
stars in the overlap areas are used to make further adjustments of
each plate's model.  This is done in an iterative approach as opposed to a
simultaneous global solution.  The procedure used consists of the following
steps:

\begin{enumerate}

\item For each plate within the plate group, identify all stars on the plate
that also have measures on other, overlapping plates.

\item Create an ``internal'' reference catalog from a weighted
average of these overlap stars' $(\alpha,\delta)$ values, obtained from
the previous iteration, for the entire plate group.  Weights are derived
from the individual positional uncertainty estimates.

\item Create an ``external'' reference catalog from the source catalog
positions of Hipparcos and Tycho2 stars present on the plate, 
adjusted for epoch using the source catalog proper motions.
External catalog positions supersede internal ones.

\item Model the plate into the combined ``internal plus external'' reference 
catalog using a general 3rd-order model
plus classical 5th-order distortion, retaining these high-order terms only
when they are significant.

\item Only after all plates have been modeled, apply the updated model
coefficients, calculating new $(\alpha,\delta)$ per object and per plate.

\item Construct an updated positional catalog of all objects on all plates 
in the group.  Use the weighted average position of multiply measured objects,
yielding a catalog with a single position per object.

\item Examine the differences in positions, comparing the newly constructed
catalog with the previous iteration's version.
If the differences are significant, go back to step 1 and
begin a new iteration.

\end{enumerate}

The presence of the Hipparcos/Tycho2 stars in the reference catalog prevents
errors from the overlap solutions from accumulating and causing a reference
system drift.  The number of iterations required for convergence was from
5 to 9 for the three plate groups.
When all is done, i.e., after sufficient convergence of the iterative
solutions, the weighted-average positions for every object on every plate
are derived and adopted as the celestial coordinates of that object, at the
weighted mean epoch for that particular star.

This procedure was applied to the first-epoch plates, the second-epoch plates,
and the second-epoch pseudo-plates that had been pasted together from the
PV CCD frame data.  For this last group, a second pasting of the CCDs
was performed using preliminary proper motions derived from a first
iteration to update all CCD data within a single pseudo-plate to the same
epoch.  Also, for the pseudo-plate regions it was realized that there were
some holes in the sky coverage due to several causes.  In areas with
single-fold coverage, inaccurate telescope pointing led to occasional
gaps between adjacent PV frames.  Additionally, some frames that had passed
a quality check at the telescope were later found to have problems that
rendered them unusable.  Finally, there were a handful of SPM fields for
which the second-epoch plates were unusable and pseudo-plates had to be created
from an incomplete number of CCD frames in these fields.  
In order to avoid having holes or cracks in the SPM4 sky
coverage for want of second-epoch positions in these cases, it was decided to
supplement the pseudo-plate fields with second-epoch positions taken from
the master input catalog.  The additional stars and galaxies added were those
with input catalog $V$ estimates less than 17.5, in general, but a cutoff of
V=16.5 was used in two galactic plane fields.  Of course, in order to appear
in the final SPM4 catalog, a corresponding detection and measure of the
object in the first-epoch plate material must exist.  Objects with proper
motions derived in this manner can be identified in the catalog; their values
of np and nc, the number of second-epoch plate and CCD measures per object,
will both be zero.

Upon completion, first-epoch positions and second-epoch positions on the
system of the ICRS had been obtained for all detected objects in the 660 fields.
Uncertainties in the positions were derived from the scatter of
multiply measured stars as a function of magnitude and this empirical relation
then adopted for all objects, whether multiply measured or not.
The positions and uncertainties were then combined to yield proper motions
and proper-motion uncertainties in a straightforward manner.  While in
theory these proper motions should be on the system of the ICRS via Hipparcos
and Tycho2, and thus absolute, in practice an additional correction is
needed.  Examining the measured proper motions of galaxies within each
field as well as the differences with Hipparcos proper motions at the bright
end, it was apparent that a residual magnitude equation remained in the
derived proper motions.  
The source of this magnitude equation is not certain.
The offsets vary with position, in particular being highly correlated
with right ascension.
Also, the offsets at the faint end, i.e., with respect to galaxies, are
much larger than those at the bright, Hipparcos-linked end.
This suggests that the cause might be a magnitude-dependent systematic in the
Tycho2 reference system used, 
one that is coherent over large angles in the
sky and is, presumably, contained in the Tycho2 proper motions.

Whatever its cause, it was decided to calculate a final correction to
absolute proper motion, per SPM field.
The form of the adopted correction is linear with magnitude, 
being anchored at the 
mean magnitude of galaxies and that of Hipparcos stars and based on the mean
proper-motion offsets of those two sets of proper-motion reference objects.  
Such a
linear correction with magnitude was derived for all 660 fields.  
The size of the corrections was less than $\sim$ 1 mas~yr$^{-1}$ at
$V \sim 8.5$ (Hipparcos magnitudes) but was typically several mas~yr$^{-1}$
at $V \sim 17.5$ (galaxy magnitudes).
More precisely, the mean proper-motion zero-point corrections
in ($\mu_{\alpha}$cos$(\delta)$, $\mu_{\delta}$) for the
660 SPM fields was (-0.22, +0.24) mas~yr$^{-1}$ at the bright end
and (-0.36, -1.83) mas~yr$^{-1}$ at the faint end, and the standard
deviations of the distribution of corrections was (0.41, 0.68) and
(0.97, 1.50) mas~yr$^{-1}$.
The actual proper-motion
correction applied to each star in the catalog was the weighted mean of the
corrections for the three closest field centers to the star, weighted by the
inverse square of the distance from the field centers.


\subsection{Photometric Reduction}

The $B,V$ photometry in the SPM4 is extremely heterogeneous.  In some cases,
it is derived from second-epoch SPM blue and visual filtered CCD frames.  
In some cases,
it is derived from the PMM measures of SPM first-epoch plates.  And in the
cases where neither of these are available or reliable, it is propagated
from the input master catalog.  In this latter group one can find relatively
good photometry from Tycho2 or less reliable extrapolations to $B$ and $V$
magnitudes from 2MASS $(J,H,K)$.  As such, it is difficult to estimate
uncertainties for much of the $B,V$ photometry listed.  SPM4 magnitude errors
are as likely to be caused by spurious radius measures or inappropriate
extrapolations as they are by signal-to-noise considerations.  Thus, we
do not provide individual uncertainty estimates for the $B,V$ photometry
listed.  We do, however, indicate the source of the $B$ and $V$ values given,
be they CCD-based, plate-based, or input catalog values.

For the purpose of identifying which image orders should be expected 
(and thus searched for)
within the list of detections on a plate or CCD, a magnitude estimate of
each star in the input master catalog is needed.  For Hipparcos and Tycho2
stars, the $B_{Tycho}$ and $V_{Tycho}$ values 
(corrected to the Johnson system) served this purpose.  
For almost all other stars, $B$ and $V$ photometry was not
available so an approximate extrapolation was derived based on 2MASS $(J,H,K)$.
Hipparcos and Tycho2 stars, which have both $B,V$ and 2MASS photometry, were
used to calibrate each SPM field with a relation of the following form,
suggested by the work of Bilir et al.~(2008),
\be
 B-J = b_0 + b_1 (J-H) + b_2 (H-K) + b_3 J (J-H) + b_4 J (H-K),
\ee
with a similar function for $V-J$.  These were used to provide an approximate
estimate of $B$ and $V$ for stars without Tycho2 photometry.  For the small
fraction of objects without either Tycho2 or 2MASS photometry, the objects were
assumed to be faint and arbitrarily assigned the magnitude limit of the
plate or CCD on which it was expected to fall.  Again, these input master
catalog $B,V$ magnitude estimates were primarily to aid in identifying the
various image orders detected.  Only in the case that there was neither SPM
plate-based photometry nor SPM CCD-based photometry did these estimates find
their way to the final catalog.

The PV and Apogee/Alta CCD frames of the second-epoch SPM survey were reduced in
typical fashion using aperture photometry with calibration into Tycho2
$V$ and $B$ photometry (corrected to the Johnson system).  When available, these
CCD-based magnitudes are provided in the SPM4 catalog, superseding any other
magnitude estimates.
As the PV camera's field of view is larger than that of the Apogee/Alta
cameras, a substantial number of stars have CCD-based $V$ magnitude
estimates and photographic $B$ estimates.

Photographic photometry based on the parameters of the image model fits of
the PMM scan data proved to be problematic.  Among the various image model
fit parameters derived, the fit radius provided the best (although still
poor) correlation with external calibrating photometry.  For extended sources,
the radius was inappropriate.  For such objects the input master
catalog's magnitude estimate was retained instead, unless there existed
CCD-based photometry.  Also, there was a large, non-gaussian scatter between
the radius measures and calibrating photometry, indicating that at times the
radius estimate was simply erroneous.  Thus, during the SPM4 plate photometric
reduction procedure, a comparison was made between the preliminarily derived
(radius-based) magnitude and that from the input master catalog.  If these
differed by more than one magnitude, it was interpreted as evidence that one
or the other was in error.  Since we could not know which, a somewhat
expedient choice was made; the fainter of the two magnitude values was
retained, under the assumption that the steepness of the luminosity function
implies that it is more likely the star is faint.  Unfortunately, the
only relevant flag that was retained per star was whether or not it had
passed through the plate photometry portion of the pipeline, not whether the
resulting magnitude estimate was truly plate-based or a retention of the input
master catalog value.  The photographic photometry was disappointingly poor
in any event.  Thus, the only truly reliable B,V photometry in the SPM4
catalog is that flagged as being CCD-based, i.e., with ib=3 and/or iv=3.

\subsection{Compilation and Content}

The final astrometry and photometry were combined with other relevant 
metrics from the reduction pipeline
to construct the SPM4 catalog.
The other quantities listed in the main catalog include: uncertainty estimates
for the astrometry; mean epochs of observation; the numbers of observations
at each epoch; an SPM4 identifier; and flags identifying the input source
catalog, source of the photometry, and whether or not the entry is matched
to an extended source or extragalactic object.
Additionally, for convenience, $J, H, K$ photometry from the 2MASS catalog
is included for all objects with a 2MASS counterpart.
Table 1 describes the contents of a single record of the main catalog,
detailing its structure and format.

The astrometric reduction procedures described earlier included checks for
spurious measures.  Typically, more than one image of an object was
available to use and, thus, eliminating a single image did not generally
mean the elimination of a star.  This applies within each of the single-epoch
positional catalogs that were generated in the process of building the SPM4.
However, there was no such automatic check during the subsequent
construction of the proper-motion catalog.  That is, there was no
post-construction filtering of the measured proper motion.  Thus, there are
undoubtedly spurious absolute proper motions in the SPM4.  The user is
cautioned that the catalog is best suited for statistical studies of
samples of stars, rather than searching for individual proper-motion
outliers within a population.  The catalog is still worthwhile for
this latter purpose,
but the user is advised to verify unusual individual proper motions
with other sources, if at all possible.  Somewhat related, it should also
be noted that the first-epoch positional catalog generated from the SPM
material was used as an early-epoch catalog in the construction of the
UCAC3 catalog, (Zacharias et al.~2010).  Thus, it is possible that some
spurious proper motions might be common to these two catalogs.

\section{Catalog Properties}

The SPM4 contains 103,319,647 stars, galaxies and QSOs.
Table 2 breaks this number down by type of second-epoch material, source
of $B$ and $V$ photometry, and intersection with a handful of relevant
external catalogs.
The spatial distribution 
of the catalog is
shown in Figure 3 as an Aitoff projection of ($\alpha,\delta)$.
Each dot represents a randomly selected one of every 500 catalog objects.
Background shading indicates the total extent of the first-epoch SPM plates.
The SPM4 has a northern boundary at $\delta=-20^{\circ}$, with a small
number of notches caused by the lack of first-epoch plates.
Low-density patches near the Galactic plane are primarily due to dust
and its associated extinction, but there are also areas in which image
crowdedness has spoiled the object detection and identification algorithms.
Caution should be exercised when using the catalog in such areas.

The magnitude distribution of the SPM4 is shown in Figure 4, based on the
same randomly selected subset of the catalog displayed in Figure 3.
An artificial bump at bright $V$ values (dotted curve) is due to
a small fraction of faint stars that suffer simultaneously from having
anomalous 2MASS colors (affecting the source
catalog $V$ extrapolation) {\it and} 
wildly miscalculated SPM photographic photometry indices, including a likely
misidentification of the image's exposure system.
This feature is a direct result of the inadequacies in the 
photographic photometry that were discussed in Sect.~4.3.
Fortunately, by filtering out objects with unrealistic $(V-J)$ values, 
these spurious stars
can be expunged, resulting in the solid
black curve shown in Figure 4.
Alternatively, the catalog's true $V$ distribution can be derived from
that portion of the SPM4 with second-epoch CCD photometry,
as the CCD-based $V$ estimates are immune from the effect.
The red curve in Figure 4 shows only the CCD-based portion of the sample.
These logarithmic curves turn over at approximately $V$=17.5 and this can
be considered the completeness limit of the catalog.
Note that the faint end of the SPM4 is determined by the 2MASS portion
of the input catalog.
As such, blue stars near the limit of $V$=17.5 may not be present in the
near-infrared 2MASS catalog and, thus, will not be included in the SPM4
even though these stars would be measurable in the SPM material.

The SPM4's single-coordinate positional uncertainty at mean epoch is
plotted as a function of magnitude in the top panel of Figure 5, based
on the one-in-500 subset.
Again we see the presence of stars with grossly underestimated $V$ magnitudes
in the dotted curves that represent the unfiltered sample.
The solid curves -- black for the overall sample, blue for the photographic
portion, and red for the CCD-based portion -- are of the sample after 
$(V-J)$ cleaning and
better represent the true run of positional uncertainty with magnitude.
The bottom panel shows the distribution in mean epoch for the sample,
broken down by second-epoch observation type, 
i.e., photographic (blue) or CCD (red).
The uncertainty-based weighting used to determine the mean epoch
accounts for the values close to the epoch extremes, especially
for objects whose second epoch is that of a CCD observation.
Note that the catalog also lists separately the mean first epoch and 
mean second epoch of observation for each object.
The insert in Figure 5 shows the distribution in epoch difference; a 
decidedly bimodal distribution, as is to be expected.

The primary emphasis of the SPM program is proper motions.
Figure 6 shows the estimated proper-motion uncertainties as a function
of magnitude for the SPM4.
In the bottom panel,
the median relation is indicated for the catalog as a whole (heavy black curve)
as well as separately for that portion of the catalog with second-epoch
plate material (blue curve) and that with second-epoch CCD measures (red curve).
As can be seen, the formal random errors in proper motion are 2 mas~yr$^{-1}$
or better over a substantial magnitude range, roughly $9<V<14$.
Beyond this, the fewer number of images per star and lower signal to
noise ratio steadily increase the errors, reaching $\sim$5 mas~yr$^{-1}$ at
$V=16.5$ and $V$=17.5 for regions with second-epoch plate and CCD
measures, respectively.

As a check on the reliability of the SPM4 proper-motion uncertainties,
at least for the portion based on second-epoch plates,
an object-by-object comparison was made with the SPM2 catalog.
The SPM2 covers a declination band from -25$^{\circ}$ to -40$^{\circ}$
and is based entirely on photographic plate material measured with the
Yale PDS microdensitometer.
Relying essentially on the same plate material, the SPM2 and SPM4 are
not independent.
However, the difference in measuring machines and centering algorithms
used, as well as significant differences in the processing procedures,
should result in individual proper motions with largely independent 
realizations of random errors.
As for the reliability of the SPM2 proper-motion uncertainties, they were
derived in a straightforward, empirical manner; by comparing proper motions 
calculated from separate blue- and yellow-plate pairs.

Approximately 300,000 stars and galaxies are in common between the SPM2 and
SPM4.
The intersection is not complete to any specific magnitude because of the
selection criteria used to construct the SPM2, 
(see Figure 4), 
but it does sample the entire magnitude
range of SPM4, with roughly half of the sample being brighter than $V=15$.
For the purpose of this analysis, the sample was divided, by right ascension,
into 30 subsamples of roughly 10,000 objects each.
Within each subsample,
simple proper-motion differences between SPM2 and SPM4 were 
normalized by their expected amplitude based on the quadrature sum of 
uncertainties from the two catalogs.
Each distribution of tentatively normalized differences was then analyzed 
using the probability plot method of Hamaker (1978) to see if it was indeed
normal.
The inner 98 percent of each subsample was used in the probability plot
analysis, i.e., an outlier frequency of just 2 percent was assumed.
The results are shown in Figure 7.
Each point shows the measured Gaussian width for a subsample's ``normalized''
distribution of proper motion errors, $\delta$ vs $\alpha$ components.
A dispersion of 1 would indicate properly calibrated uncertainty estimates.
The average value of the subsamples' measured dispersion was 0.95 and 0.97,
in the two components.
This implies that the estimated uncertainties are well-calibrated, to within
a few percent.

Evaluating the systematic accuracy of the SPM4 proper motions is more
challenging.
Recall that a final zero-point correction was applied, one that was
linear with magnitude
and tied to Hipparcos proper motions on the bright end and external galaxies
and QSOs on the faint.
Therefore, direct comparisons at this stage to either Hipparcos proper motions
or to extragalactic sources would not be meaningful.
Other possible proper-motion catalogs that might be considered for comparison
are the UCAC2 and Tycho2.
The accuracy of the UCAC2 is similar to its precision, on the order of a 
few mas~yr$^{-1}$, and therefore would not be useful for this purpose.
A comparison to Tycho2 would only reflect the systematics found in the
preliminary SPM4 proper motions discussed in Section 4.2 and attributed
to magnitude effects in the Tycho2 proper motions that had been utilized
in setting the reference system for our first-epoch positions.
Comparison with deeper proper-motion surveys, for instance those based on
first-epoch astrometry from the USNO-A2 or USNO-B1 catalogs or other
Schmidt survey catalogs would also be fruitless, as will be illustrated by 
an example presented in Section 7.

Instead, we require systems of stars (in order to beat down the random
errors) with previously determined absolute proper motions.
We have chosen five globular clusters with well-determined absolute
proper motions and now redetermine their proper motions from SPM4,
for comparison.
Table 3 lists the five clusters examined.
Reference values of absolute proper motion and their associated
uncertainties are taken from the literature.
For each cluster, 
stars are selected from the SPM4 catalog by manually trimming in
position, proper-motion, and color-magnitude space.
The numbers of SPM4 stars per cluster varied from a few hundred to just over
a thousand.
The mean motion of each cluster was determined, as was the contribution
to the uncertainty of the mean from random catalog errors, (listed in
parentheses in Table 3).
Differences between the SPM4 determinations and the literature values were
then scaled by the expected amplitude of deviation, i.e., the quadrature sum of
the literature uncertainty, the contribution from the SPM4 random errors, and
an SPM4 systematic contribution to be solved for.
For this rather small sample of five test objects, an SPM4 systematic
uncertainty of 1.7 mas~yr$^{-1}$ is needed to make the distribution of
scaled differences normal.
This is larger than the anticipated accuracy of 1 mas~yr$^{-1}$ expected
from our previous experience with PMM measures of SPM material.
We suspect the crowded images within clusters do not lend themselves
well to the survey-mode pipelining necessary to build the SPM4.
Another possibility is that the linear magnitude correction applied near
the end of the SPM4 reductions may, in some fields, be harmful.
For instance, if we use the Hipparcos stars in the neighborhood of our
test clusters to make a constant correction to the SPM4-determined absolute
motions, the systematic uncertainty required to normalize
the distribution of scaled proper-motion differences drops from 1.7
mas~yr$^{-1}$ to 1.3 mas~yr$^{-1}$.

Finally, as discussed in Section 4.3, the $(B,V)$ photometry listed in the
SPM4 catalog is extremely heterogeneous, being a mixture of CCD photometry,
photographic photometry based on the PMM-measured image radii, and 
extrapolations of 2MASS $(J,H,K)$.
The photographic photometry suffers not only from Gaussian errors associated
with the plate scanning and image fitting process but there is also a
subset of images with wildly erroneous magnitude estimates, estimates that
may be off by five magnitudes or more.
These are due to
misidentification of image systems and/or inappropriate extrapolations
for stars with poorly measured 2MASS colors.
Thus, while the catalog does not list photometric uncertainty estimates,
flags are included that give the source of the $(B,V)$ data.
Only the CCD photometry should be trusted.
The precision of the CCD-based $V$ data is $\sim$0.02 mag for $V < 15$,
rising sharply to $\sim 0.09$ mag at $V$=18.
The blue camera produces inferior images but the exposures go deeper.
The result is CCD-based $B$ uncertainties that are $\sim$0.03 for stars
brighter than $B$=14, rising slowly to $\sim$0.05 at $B$=19.
Conversely, the standard deviation of the photographic photometry is
estimated to be on the
order of 0.5 magnitudes, but the error distribution has a long, non-Gaussian
tail toward artificially bright magnitudes. 
The lack of reliability of the $(B,V)$ photometry was a compelling reason
for including 2MASS $(J,H,K)$ photometry in the SPM4.

\section{Catalog Versions}

Several previous versions of SPM catalog have been released and 
remain available.
These should not necessarily be considered superseded by the current SPM4.
A brief description of the major SPM catalog releases and their differences
is warranted, 
as their relative strengths and weaknesses
are relevant to the potential user.

The SPM1 is based on Yale PDS measures of plates from 30 SPM fields around
the South Galactic Pole.
It covers $\sim$720 sq deg and contains 58,887 objects.
Two subversions were released, the SPM 1.0 and SPM 1.1, the latter being
superior in that it
included improvements made to the magnitude-correction scheme.

The SPM2 is also based on PDS measures but covers a larger swath of sky;
156 SPM fields totaling $\sim$3700 sq deg.
These are all fields in the -25$^{\circ}$ to -40$^{\circ}$ declination bands,
but minus the low galactic latitude fields.
The SPM2, (the one and only release being the SPM 2.0), contains 321,608
objects.
As with the SPM1, the SPM2 is not intended to be complete to any magnitude
and, indeed, is far from it.
The slow throughput of the PDS microdensitometer precludes measuring all
of the images present in the plate material.
Thus, an input catalog was adopted, composed of objects of interest gleaned
from the literature, utility objects (i.e., images necessary for the reduction
procedures), visually confirmed galaxies, and a large number of randomly
selected anonymous stars intended for kinematic investigations.
The result is a magnitude distribution for the SPM2 that is skewed strongly
toward brighter stars.
This is illustrated in the insert panel of Figure 4.

Absolute proper motions in both the SPM1 and SPM2 are referred to galaxies.
Positions are on the system of the ICRS via a subset of Tycho2 stars.
The SPM1 and SPM2 should be read with care, as the positions are given
at the epoch of the Tycho2 catalog, 1991.25.
Subsequent versions, SPM3 and SPM4, are presented at the more 
conventional epoch of 2000.

The SPM3 consists of the same 156 fields comprising the SPM2, but in the
case of the SPM3 the plates were scanned using the PMM of the Naval
Observatory Flagstaff Station (NOFS).
Image detections and centers were those of the PMM scanning/measurement
pipeline.
For convenience, an input catalog was still utilized in transforming these
into an SPM 3.0 catalog, but now an attempt was made to have the input 
catalog be complete to roughly the limit of the SPM plate material.
This was accomplished by combining source catalogs consisting of Hipparcos,
Tycho2, UCAC2 and a magnitude-clipped USNO-A2.
Shortly thereafter the 2MASS all-sky catalog was released and the USNO-A2
portion of our input catalog was replaced with 2MASS resulting in version
SPM 3.1.
Two subsequent improvements (fixes, actually) in the reduction code led to
the definitive SPM3 version, SPM 3.3.
The SPM 3.3 catalog contains $\sim$11 million stars and galaxies.
It differs from the SPM1 and SPM2 catalogs not only in magnitude completeness,
but also in the absolute proper motion reference used.
The SPM 3.3 proper motions are tied to the ICRS entirely through the
use of Hipparcos stars.

The current catalog, SPM4, is the first release that combines PMM measures
of SPM photographic plates with second-epoch CCD measures.
All previous SPM catalogs were based exclusively on photographic material.
The SPM4 uses a slightly expanded input catalog source list, compared to
that of the SPM3, as described in
Sect.~4.
It also differs from previous versions in the manner in which the proper
motions are tied to an inertial frame, using Hipparcos stars on the bright
end and galaxies and QSOs on the faint end, with a linear interpolation as
a function of magnitude.
Other important differences with previous SPM catalogs, such as the creation of
separate first- and second-epoch positional catalogs during its construction,
can be found in the detailed description of Sect.~4.

Which catalog to choose depends on the intended use.
The incomplete SPM2 is based on PDS measures, which are more precise than
PMM measures of the same plates.
Also, the SPM2 was a small enough project that a significant amount of
human interaction could be made during the reduction and evaluation process,
on a field-by-field basis.
Thus, the SPM2 is less likely to contain individual stars or entire fields 
that are affected by undetected ``blunders.''
The sheer numbers involved in the SPM3 and SPM4 precluded a high level
of human checking of the results.
On the other hand, the severe incompleteness of the pre-SPM3 catalogs 
places limitations on their usefulness.
Finally, if a star or set of stars of interest are in both the SPM3 and
SPM4, the SPM3 astrometry is likely to be marginally less precise, due to the
refinements in centering implemented for the SPM4.  
However, systematics
in the absolute proper motions of the SPM3 are expected to be smaller
than those in the SPM4.
This is because SPM2 astrometry was used to help calibrate
the SPM3, while the SPM4, which extends well beyond the coverage of the SPM2,
could not benefit from such a calibration.

\section{Summary}

We present a comprehensive description of the SPM4 catalog, 
the latest and largest
astrometric product of the Yale/San Juan Southern Proper Motion program.
The SPM4 contains positions, absolute proper motions, and $B,V$ photometry
for over 100 million stars and galaxies south of -20$^{\circ}$ declination.
Emphasis is on the astrometric aspect of the catalog, particularly the
absolute proper motions, which have a precision of $\sim$2 mas~yr$^{-1}$
for well-measured stars.
The proper motions are on the system of the ICRS and have an estimated
systematic accuracy on the order of 1 mas~yr$^{-1}$.

It is worth noting that the SPM4 is based on an astrometric observing 
program that was intended from its outset to generate absolute proper motions
of high precision and accuracy.
This sets it apart from recent proper-motion catalogs that rely on 
photographic Schmidt survey measures for early-epoch positions.
The distinction in precision is readily seen by comparing SPM4 proper motions
with, for instance, those from the XPM catalog (Fedorov et al.~2010)
and the PPMXL catalog (Roeser et al.~2010).
Figure 8 shows such a comparison for
a 30 arcmin x 30 arcmin region centered on the globular cluster NGC 6397.
The intrinsic proper-motion dispersion of the cluster is negligible and that of
the field is relatively low at this Galactic pointing
($l=338^{\circ}, b=-12^{\circ}$).
Only the SPM4 proper motions show a discernible separation of cluster and
field stars.

The SPM4 is well-suited for kinematic studies of the Galaxy, particularly
when combined with complementary radial-velocity data.
An example of such a 3-D study is that of Casetti-Dinescu et al.~(2011) in
which the SPM4 is merged with the second data release of RAVE, the RAdial 
Velocity Experiment survey (Zwitter et al.~2008).
By isolating red clump stars common to the two surveys, Casetti-Dinescu
et al.~are able to derive the properties of the thick-disk velocity
ellipsoid; including its dispersions and tilt angles, as well as the vertical 
gradient of the rotational component.
They have also used this sample to deduce an interesting disparity between 
the currently accepted value for the motion of the local standard of rest
based on stars within 1 kpc and the motion observed relative to more distant
tracers, beyond 1 kpc.
Other intriguing results await further exploitation of the SPM4 data.

The entire SPM4 catalog is available via ftp 
download\footnote{see www.astro.yale.edu/astrom/} or, 
upon request to the first author,
in compressed form on a single DVD disc.
We are also in the process of making the SPM4 available via the web-based VIZIER
service, facilitating those users who wish to make extractions around areas of 
interest.


The SPM program is a decades-long endeavor involving the
participation of several institutions and countless people.  
These ``countless'' many, beyond the authors of this manuscript, 
have contributed greatly
to the success of the SPM program and its culmination 
with the release of the SPM4 catalog.
The following
is a modest attempt at listing all those involved:
Imants Platais {\it (Johns Hopkins)},
Vera Kozhurina-Platais {\it (STScI)},
Reed Meyer {\it (TripAdvisor LLC, Boston)},
Arnold Klemola {\it (Lick Obs.)},
Ren\'{e} M\'{e}ndez {\it (Univ.~de Chile)},
Xinjian Guo {\it (Yale)},
Paulo Holvorcem {\it (Univ.~Estadual de Campinas, Brazil)},
John T. Lee {\it (Interactive Data, Boxborough, Mass.)},
Zhenghong Tang {\it (Shanghai Astronomical Obs.)},
Vladimir Korchagin {\it (Federal Southern Univ., Russia)},
Ting-Gao Yang {\it (Chinese Academy of Sciences, Time Service Center)},
Wen-Zhang Ma {\it (Beijing Normal Univ.)},
Chun-lin Lu {\it (Purple Mountain Obs.)}
Gary Wycoff {\it (USNO)},
Charlie Finch {\it (USNO)},
Jin-Fuw Lee {\it (IBM, deceased)}.

We are grateful to the National Science Foundation for their substantial 
support in the form of a series of grants spanning more than four decades,
the most recent being grant AST04-0908996;
the University of San Juan for extensive logistical and personnel support
throughout the course of the survey; the Argentine CONICET for funding of
some of the instrumentation; and Yale University for critical financial 
support during the completion of the SPM program.  
Also, we are thankful for partial support from the NSF 
under grants
PHY 02-16783 and PHY 08-22648 (Physics Frontier Center/Joint Institute 
for Nuclear Astrophysics).
The SPM program would 
not have begun were it not for an initial grant from the Ford Foundation, 
which we also gratefully acknowledge.  Finally, we are indebted to our 
observers who provided the raw material upon which this catalog is based.

This publication makes use of data products from the Two Micron All Sky Survey, 
which is a joint project of the University of Massachusetts and the Infrared 
Processing and Analysis Center/California Institute of Technology, funded by 
NASA and the NSF.

\clearpage

\begin{table}
\begin{center}
\begin{tabular}{rrllll}
\multicolumn{6}{l}{Table 1.  SPM4 contents} \\
\tableline
\tableline
\vspace*{0.2cm}
col & bytes & format & name & unit & description \\
\tableline
  1& 001-012 & f12.7 & ra  & deg   & right ascension (ICRS, epoch=2000.0) \\
  2& 013-024 & f12.7 & dec & deg   & declination     (ICRS, epoch=2000.0) \\
  3& 025-030 & f6.1  & era & mas   & uncertainty in right ascension \\
  4& 031-036 & f6.1  & edec & mas   & uncertainty in declination \\
  5& 037-045 & f9.2  & pma & mas/yr & absolute proper motion in ra ($\mu_{\alpha}$cos$\delta$) \\
  6& 046-054 & f9.2  & pmd & mas/yr & absolute proper motion in dec \\
  7& 055-061 & f7.2  & epma & mas/yr & uncertainty in pma \\
  8& 062-068 & f7.2  & epmd & mas/yr & uncertainty in pmd \\
  9& 069-074 & f6.2  & B   & mag   & B magnitude \\
 10& 075-080 & f6.2  & V   & mag   & V magnitude \\
 11& 081-083 & i2,i1 & ib,iv  &       & source flags for B and V magnitudes \\
 12& 084-089 & f6.2  & epav & yrs   & weighted mean epoch of observation minus 1950 \\
 13& 090-095 & f6.2  & ep1 & yrs   & weighted mean 1st epoch minus 1950 \\
 14& 096-101 & f6.2  & ep2 & yrs   & weighted mean 2nd epoch minus 1950 \\
 15& 102-104 & i3    & mp  &       & number of 1st-epoch plates used \\
 16& 105-106 & i2    & np  &       & number of 2nd-epoch plates used \\
 17& 107-108 & i2    & nc  &       & number of 2nd-epoch CCD frames used \\
 18& 109-119 & i11.10 & spmid &    & spm4 identifier = field no. + input catalog line \\
 19& 120-121 & i2    & igal  &    & galaxy/extended-object flag \\
 20& 122-122 & i1    & icat  &    & input catalog source flag \\
 21& 123-129 & f7.3  & J     &    & 2MASS J magnitude \\
 22& 130-136 & f7.3  & H     &    & 2MASS H magnitude \\
 23& 137-143 & f7.3  & K     &    & 2MASS K magnitude \\
\tableline
\\
\multicolumn{6}{l}{Additional notes by column;} \\
    3,4.& \multicolumn{5}{l}{Positional errors are given at the mean epoch of
observation, i.e., column 13.} \\
    11.& \multicolumn{5}{l}{If ib/iv is (1; 2; 3), B/V is from (input catalog;
1st-epoch plates; 2nd-epoch CCD).} \\
     14.& \multicolumn{5}{l}{Preliminary proper motions were used to put
all CCD ``pseudo-plates'' at epoch 2007.0.} \\
    17. & \multicolumn{5}{l}{If the number of frames is greater than 9, nc is set to 9.} \\
    18. & \multicolumn{5}{l}{The first three digits indicate the SPM field
number.  The final seven digits are the} \\ 
        & \multicolumn{5}{l}{running number from that field's input master list,
a concatenation of the input catalogs.} \\
    19. & \multicolumn{5}{l}{If 0, there is no indication the object is non-stellar.} \\
        & \multicolumn{5}{l}{If (1; 2; 3), then (2MASS extended source;
LEDA galaxy; Veron-Cetty \& Veron QSO).} \\
    20. & \multicolumn{5}{l}{If (1; 2; 3), the object is from (Hipparcos; 
Tycho2; UCAC2).} \\
        & \multicolumn{5}{l}{If 4, the object is from the 2MASS point source catalog (psc).} \\
        & \multicolumn{5}{l}{If 5, the object is from the 2MASS extended source catalog (xsc).} \\
        & \multicolumn{5}{l}{If 6, the object is from the LEDA galaxy catalog.} \\
        & \multicolumn{5}{l}{If 7, the object is from the Veron-Cetty \& Veron QSO catalog).} \\
       & \multicolumn{5}{l}{For objects in multiple catalogs, the first (lowest numbered) catalog source is given.} \\
 21-23.& \multicolumn{5}{l}{J, H, K from 2MASS (psc) for matched objects, otherwise = 0.0.} \\

\end{tabular}
\end{center}
\end{table}

\begin{table}
\begin{center}
\begin{tabular}{lr}
\multicolumn{2}{l}{Table 2.  SPM4 statistics.} \\
\tableline
\tableline
\vspace*{0.2cm}
Total number of SPM4 objects           &                   103,319,647  \\
Objects with 2nd-epoch astrometry...   &                                \\
~~~from CCD measures                   &                    65,355,419  \\
~~~from plate measures                 &                    27,042,797  \\
\vspace*{0.2cm}
~~~from source-catalog positions       &                    10,921,431  \\
Objects with $B$ photometry...  &                                \\
~~~from 2nd-epoch CCD frames           &                     8,338,220  \\
~~~from 1st-epoch plates               &                    83,295,995  \\
\vspace*{0.2cm}
~~~from source-catalog photometry      &                    11,685,432  \\
Objects with $V$ photometry...  &                                \\
~~~from 2nd-epoch CCD frames           &                    65,262,893  \\
~~~from 1st-epoch plates               &                    37,624,399  \\
\vspace*{0.2cm}
~~~from source-catalog photometry      &                       432,355  \\      
Objects in common with...  &                                       \\
~~~Hipparcos                           &                        43,889  \\
~~~SPM2                                &                       301,624  \\
~~~2MASS extended-source catalog       &                       214,279  \\
~~~LEDA galaxies catalog               &                        85,519  \\
\vspace*{0.2cm}
~~~V-C\&V quasar catalog               &                         1,341  \\
Objects {\it without} a 2MASS cross-id &                       153,192  \\
\tableline
\end{tabular}
\end{center}
\end{table}

\begin{table}
\begin{center}
\begin{tabular}{rrrcrrcrr}
\multicolumn{7}{l}{Table 3.  Globular cluster proper motions $^{a}$} \\
\tableline
\tableline
Cluster: & 
   \multicolumn{2}{c}{Reference} & &
   \multicolumn{2}{c}{SPM4} & &
   \multicolumn{2}{c}{SPM4 - Ref} \\
\multicolumn{1}{l}{NGC...} & 
  \multicolumn{1}{c}{$\mu_{\alpha}$cos$\delta$} & 
  \multicolumn{1}{c}{$\mu_{\delta}$} & &
  \multicolumn{1}{c}{$\mu_{\alpha}$cos$\delta$} & 
  \multicolumn{1}{c}{$\mu_{\delta}$} & &
  \multicolumn{1}{c}{$\Delta\mu_{\alpha}$cos$\delta$} & 
  \multicolumn{1}{c}{$\Delta\mu_{\delta}$} \\
\tableline
 104  
   & 5.83$\pm$0.17$^b$ & -2.59$\pm$0.17$^b$ & & 7.67~(0.11) & -3.30~(0.10) & &
   1.84~~ & -0.71 \\
3201 
   & 5.28$\pm$0.32$^c$ & -0.98$\pm$0.33$^c$ & & 7.18~(0.10) & -3.39~(0.10) & &
   1.90~~ & 2.41 \\
5139 
   & -5.08$\pm$0.35$^d$ & -3.57$\pm$0.34$^d$ & & -3.58~(0.13) & -4.49~(0.15) & &
   1.50~~ & -0.92 \\
6121
   & -12.76$\pm$0.23$^e$ & -19.38$\pm$0.25$^e$ & & -12.70~(0.33) & -17.40~(0.33) & &
   0.06~~ & 1.98 \\
6397
   & 3.56$\pm$0.04$^f$ & -17.34$\pm$0.04$^f$ & & 1.17~(0.06) & -15.65~(0.06) & &
   -2.39~~ & 1.69 \\
\tableline
\multicolumn{9}{l}{$^{a}$ Proper motions expressed in mas~yr$^{-1}$} \\
\multicolumn{9}{l}{$^{b}$ Anderson \& King 2003 plus
Kallivayalil et al.~2006; Freire et al.~2003: (weighted average)} \\
\multicolumn{9}{l}{$^{c}$ Casetti-Dinescu et al.~2007} \\
\multicolumn{9}{l}{$^{d}$ Dinescu et al.~1999} \\
\multicolumn{9}{l}{$^{e}$ Dinescu et al.~1999, Bedin et al.~2003,
Kalirai et al.~2004: (weighted average)} \\
\multicolumn{9}{l}{$^{f}$ Milone et al.~2006, Kalirai et al.~2007: (weighted
average)} \\
\end{tabular}
\end{center}
\end{table}

\clearpage

\begin{figure} 
\centering
\includegraphics[scale=0.85,angle=-90]{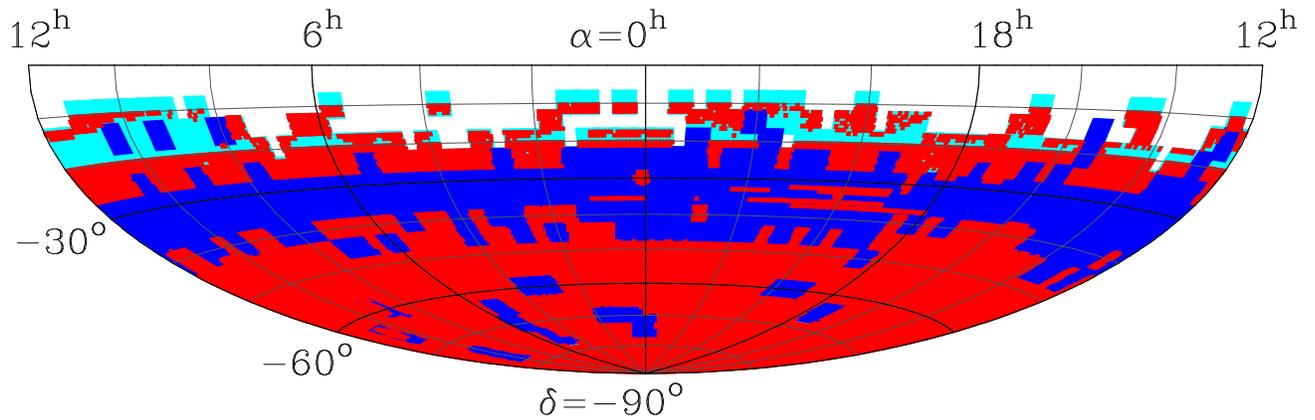}
\caption{Sky coverage of the SPM surveys, in equatorial coordinates.
The light blue area indicates first-epoch plate coverage.
Overplotted in dark blue is the area covered by second-epoch photography.
Areas with second-epoch CCD observations are shown in red.
Second-epoch CCD measures were used for roughly two thirds of the 
areal coverage of the SPM4 catalog.
}
\end{figure} 
 
\begin{figure} 
\centering
\includegraphics[scale=0.85]{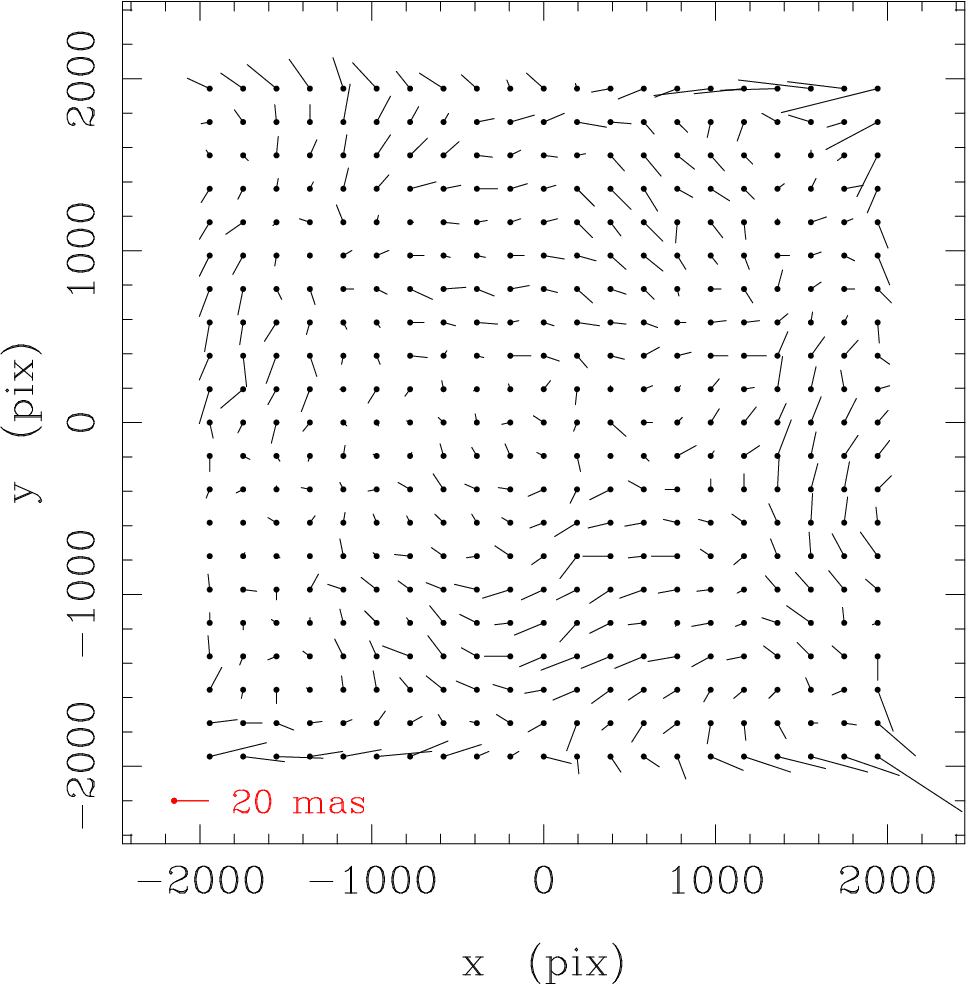}
\caption{Field-distortion correction mask for the PV CCD/filter
combination.
The mask was derived by stacking residuals of PV frames reduced
into UCAC2 positions.
The scale of the vectors is as indicated in the lower left.
}
\end{figure} 

\begin{figure}
\centering
\includegraphics[scale=0.85,angle=-90]{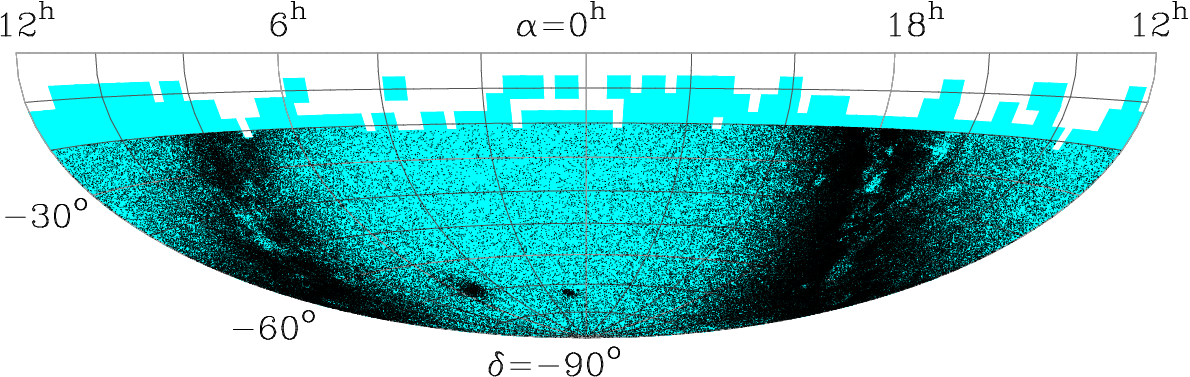}
\caption{Spatial distribution of the SPM4 catalog.
Each dot represents one in 500 objects, randomly selected from the catalog.
The background shading shows the total extent of the first-epoch SPM
photographic survey.
The northern edge of the SPM4 catalog is $\delta$=-20$^{\circ}$, minus a
handful of slight indentations where first-epoch plates are absent.
}  
\end{figure}

\begin{figure} 
\centering
\includegraphics[scale=0.85]{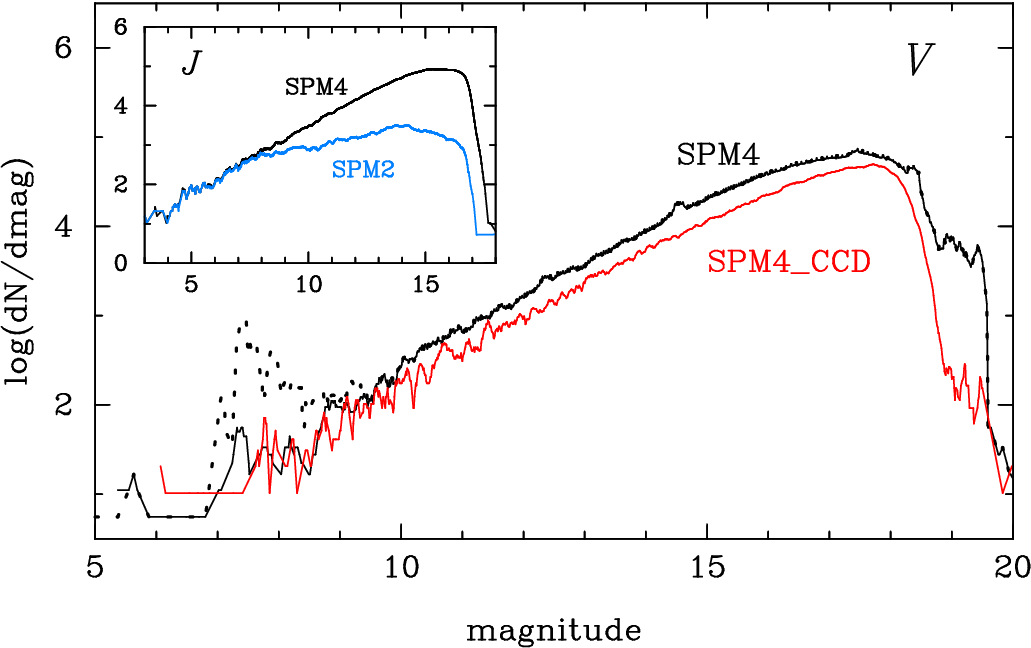}
\caption{Magnitude distribution of the SPM4.
The main panel shows the $V$ distribution of a randomly selected 0.2\% of
the catalog, consisting of both photographic and CCD photometry.
The dotted curve shows the distribution without any cleaning of spurious
$V$ estimates.
The solid curve is the result after discarding stars with unrealistic
$V-J$ values.
The red curve shows the distribution of those objects in the subsample
whose $V$ magnitude is CCD-based.
The logarithmic distribution turns over at $V \sim$ 17.5.
The insert shows a comparison of the magnitude distributions of the
SPM4 and PDS-based SPM2 catalogs.
All catalog objects within a specific mid-latitude region, 200 sq deg in area, 
are included and the
distributions are constructed as a function of 2MASS $J$ so as to avoid
spurious features in the SPM4 $V$ photographic photometry.
}
\end{figure}

\begin{figure}
\centering
\includegraphics[scale=0.85]{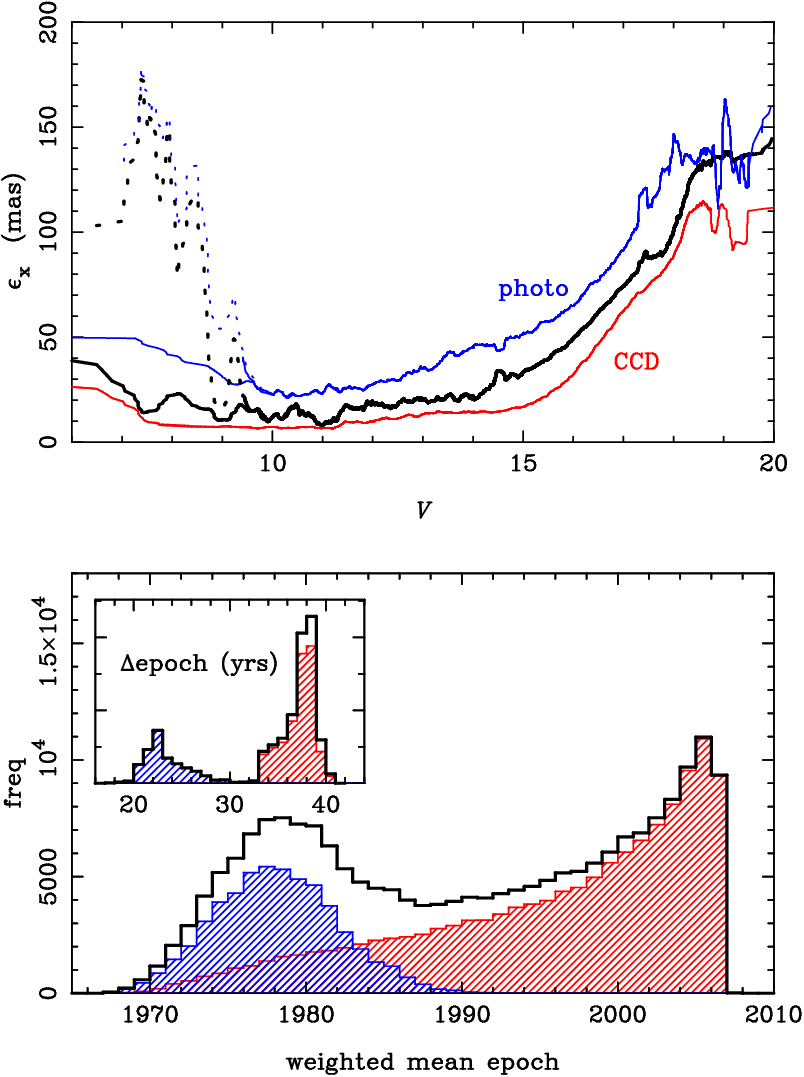}
\caption{Single-coordinate positional uncertainties at mean epoch as a function
of $V$ magnitude.
The top panel shows the median uncertainty for the catalog as a whole (black
curves), for that portion with second-epoch photography (blue curves), and
for that portion with second-epoch CCD astrometry (red curve).
A group of faint stars with incorrect photographic $V$ values accounts
for the artificial rise at $V < 10$ (dotted curves).
See the text for further explanation.
Filtering the sample by $(V-J)$ allows these erroneous stars to be expunged,
revealing the true magnitude of the uncertainties at the bright end
(solid curves).
In the bottom panel, the distribution in weighted mean epoch is shown 
(black curve)
along with the distributions of objects separated by second-epoch observation
type; photographic plates (blue hatching) or CCD (red hatching).
The uncertainty-based weighting scheme and superior astrometry of the CCD
second-epoch positions skews the distribution toward the CCD era.
The insert shows the distribution in epoch difference, using similar
color coding.
}
\end{figure}

\begin{figure}
\centering
\includegraphics[scale=0.75]{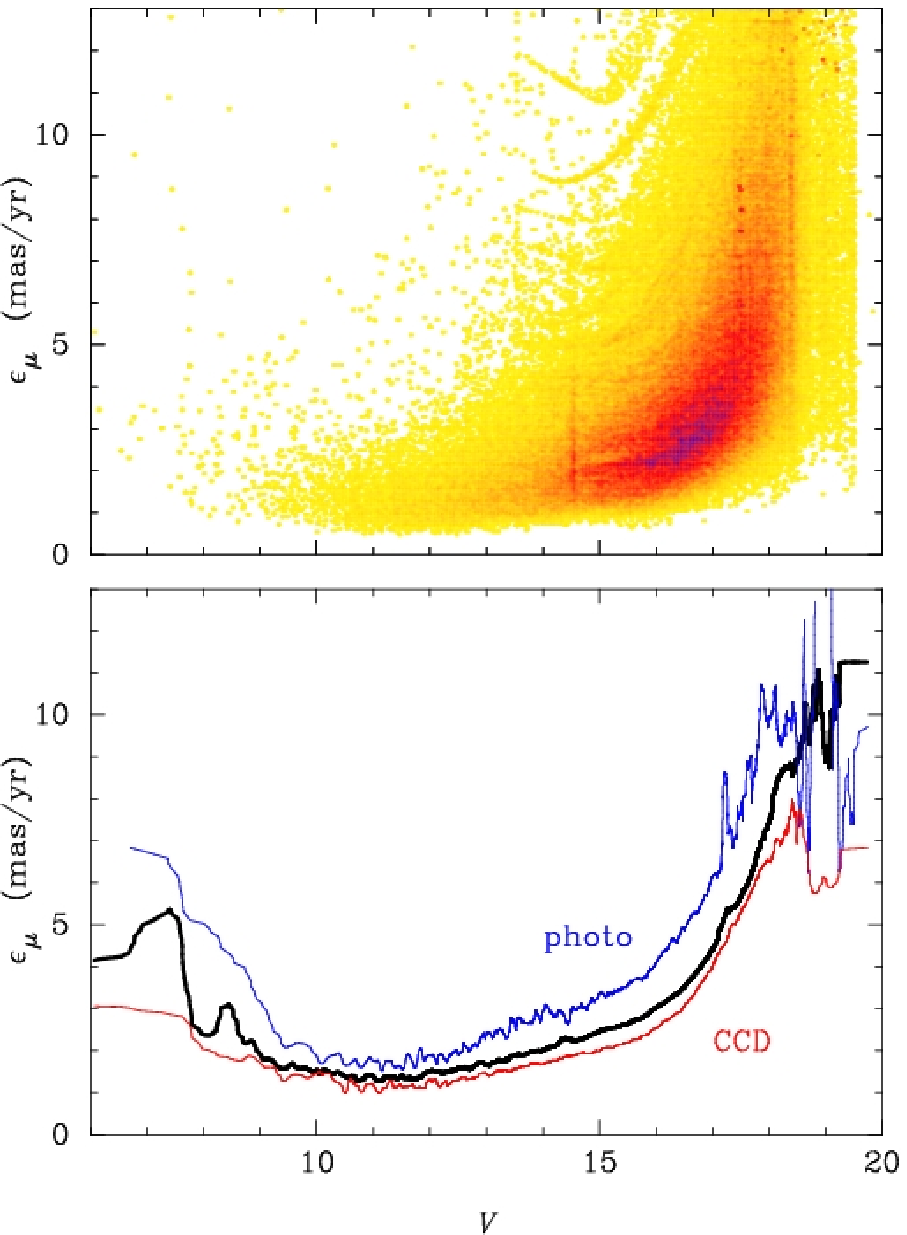}
\caption{Estimated proper-motion uncertainties as a function
of $V$ magnitude.
The top panel shows the density of SPM4 objects in the $\epsilon_{\mu}$-$V$
plane, where $\epsilon_{\mu}$ is the single-coordinate proper-motion
uncertainty.
The color coding is linear with density.
In the bottom panel, the median relation is shown for the catalog
as a whole (black curve).
Also shown are the
median relations for objects with photographic second-epoch astrometry (blue)
and those with CCD second-epoch astrometry (red).
The contaminating set of stars with erroneous photographic $V$ estimates
brighter than $\sim$10 
have been removed by trimming the sample in $(V-J)$.
}
\end{figure}

\begin{figure}
\centering
\includegraphics[scale=0.85]{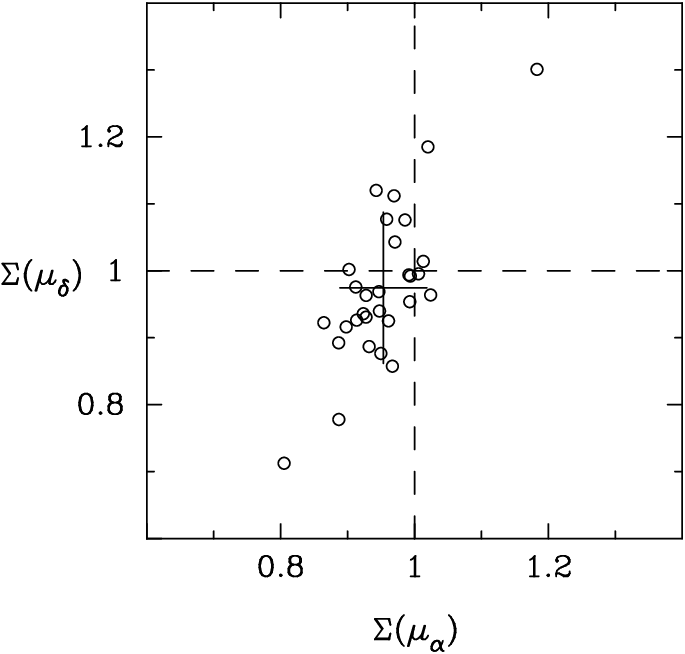} 
\caption{An evaluation of SPM4 proper-motion uncertainty estimates
by comparison to SPM2.
Each point represents $\sim$10,000 objects in common between the two
catalogs, binned by right ascension. 
Proper-motion differences, SPM4-SPM2, within each subsample are scaled by the
expected amplitude of deviation, i.e., the quadrature sum of the uncertainty 
estimates from the two catalogs.
The distributions of scaled differences within each bin are then analyzed 
using the probability plot method.
The distributions' measured Gaussian dispersions,
along $\mu_{\alpha}$ and $\mu_{\delta}$,
are plotted for the 30 subsamples.
The cross indicates the mean value of (0.95, 0.97),
which can be compared to the expected value of (1.00, 1.00) for
perfectly calibrated uncertainties.
} 
\end{figure}

\begin{figure}
\centering
\includegraphics[scale=0.85]{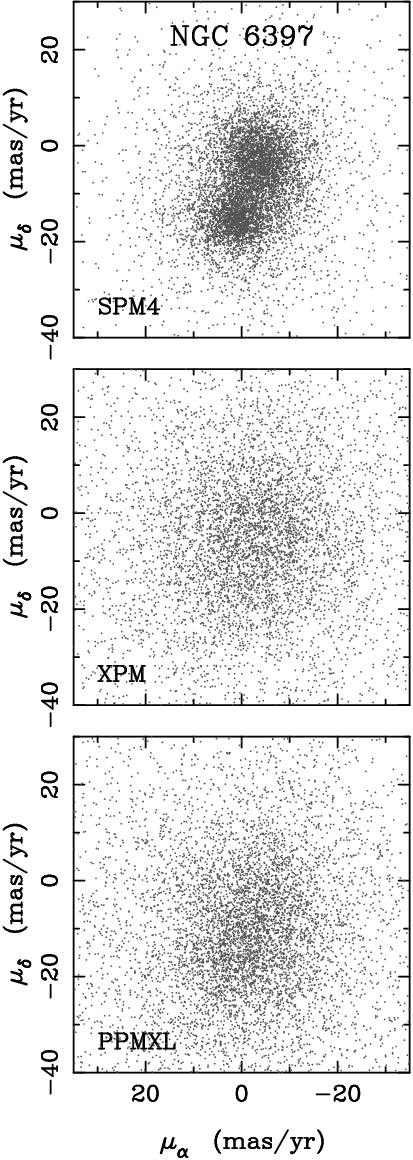}
\caption{A comparison of SPM4, XPM, and PPMXL catalogs in the field of
NGC 6397.
The proper-motion vector point diagrams are of all objects within a
30 arcmin x 30 arcmin region centered on the cluster.
The top panel shows SPM4 proper motions, 
the middle panel shows the XPM proper motions, and
the bottom panel shows PPMXL proper motions.
The field-star and cluster distributions are readily discernible only in the
SPM4 data.
}
\end{figure}

\end{document}